\documentclass[preprint2]{aastex61}

\shorttitle{Magnetic field of a coronal cavity}
\shortauthors{Chen et al.}

\begin{document}

\title{Diagnosing the magnetic field structure of a coronal cavity observed during the 2017 total solar eclipse}

\correspondingauthor{Hui Tian}
\email{huitian@pku.edu.cn}

\author{Yajie Chen}
\affil{School of Earth and Space Sciences, Peking University, Beijing 100871, China}

\author{Hui Tian}
\affiliation{School of Earth and Space Sciences, Peking University, Beijing 100871, China}

\author{Yingna Su}
\affiliation{Key Laboratory for Dark Matter and Space Science, Purple Mountain Observatory, CAS, Nanjing 210008, China}
\affiliation{School of Astronomy and Space Science, University of Science and Technology of China, Hefei, Anhui 230026, China}

\author{Zhongquan Qu}
\affiliation{Yunnan Observatories, Chinese Academy of Sciences, Kunming 650011, China}

\author{Linhua Deng}
\affiliation{Yunnan Observatories, Chinese Academy of Sciences, Kunming 650011, China}

\author{Patricia R. Jibben}
\affiliation{Harvard-Smithsonian Center for Astrophysics, 60 Garden Street, Cambridge, MA 02138, USA}

\author{Zihao Yang}
\affiliation{School of Earth and Space Sciences, Peking University, Beijing 100871, China}

\author{Jingwen Zhang}
\affiliation{School of Earth and Space Sciences, Peking University, Beijing 100871, China}

\author{Tanmoy Samanta}
\affiliation{School of Earth and Space Sciences, Peking University, Beijing 100871, China}

\author{Jiansen He}
\affiliation{School of Earth and Space Sciences, Peking University, Beijing 100871, China}

\author{Linghua Wang}
\affiliation{School of Earth and Space Sciences, Peking University, Beijing 100871, China}

\author{Yingjie Zhu}
\affiliation{School of Earth and Space Sciences, Peking University, Beijing 100871, China}

\author{Yue Zhong}
\affiliation{Yunnan Observatories, Chinese Academy of Sciences, Kunming 650011, China}

\author{Yu Liang}
\affiliation{Yunnan Observatories, Chinese Academy of Sciences, Kunming 650011, China}

\begin{abstract}

We present an investigation of a coronal cavity observed above the western limb in the coronal red line Fe X 6374 {\AA} using a telescope of Peking University and in the green line Fe XIV 5303 {\AA} using a telescope of Yunnan Observatories, Chinese Academy of Sciences during the total solar eclipse on 2017 August 21. A series of magnetic field models are constructed based on the magnetograms taken by the Helioseismic and Magnetic Imager onboard the Solar Dynamics Observatory (SDO) one week before the eclipse. The model field lines are then compared with coronal structures seen in images taken by the Atmospheric Imaging Assembly on board SDO and in our coronal red line images. The best-fit model consists of a flux rope with a twist angle of 3.1$\pi$, which is consistent with the most probable value of the total twist angle of interplanetary flux ropes observed at 1 AU. Linear polarization of the Fe XIII 10747 {\AA} line calculated from this model shows a ``lagomorphic" signature that is also observed by the Coronal Multichannel Polarimeter of the High Altitude Observatory. We also find a ring-shaped structure in the line-of-sight velocity of Fe XIII 10747 {\AA}, which implies hot plasma flows along a helical magnetic field structure, in the cavity. These results suggest that the magnetic structure of the cavity is a highly twisted flux rope, which may erupt eventually. The temperature structure of the cavity has also been investigated using the intensity ratio of Fe XIII 10747 {\AA} and Fe X 6374 {\AA}. 

\end{abstract}


\section{Introduction} \label{sec:intro}
Solar coronal cavities are elliptical regions of reduced coronal emission with rarefied density \citep{fuller2009, gibson2010, forland2013}, which are usually observed to be associated with prominences above the solar limb \citep{Tandberghanssen1995, hudson1999}. The first records of cavities were obtained in white light (WL) during the solar eclipse on 1898 January 22 \citep{Wesley1927}. Since then cavities have been observed in different wavelength ranges such as in radio, WL, extreme ultraviolet (EUV) and soft X-ray \citep[e.g.,][]{Marque2002, Marque2004, gibson2006, Regnier2011, hudson1999, hudson2000, Heinzel2008}. \citet{Reeves2012} found elevated temperatures in the core of a cavity. A cavity may erupt and be observed as part of Coronal Mass Ejections (CMEs) \citep[e.g.,][]{illing1986, Tandberghanssen1995, hudson1999}. However, quiescent cavities can exist for days or weeks in equilibrium before eruption \citep{gibson2006}. Investigations of the magnetic structures of quiescent cavities are important to establish the pre-eruption magnetic configurations and understand magnetic structures of the corona as a whole. 

However, precise magnetic field measurements through the Zeeman effect are limited to the photosphere. Direct measurement of the coronal magnetic fields is difficult.This is because the Zeeman splitting in corona is often negligible due to the weak magnetic field, previously estimated to be of the order of 10 G \citep{Harvey1969,Kuhn1995,lin2000}, and most coronal spectral lines have a very large broadening due to the high temperature. \citet{lin2000,lin2004} have tried to measure the coronal magnetic fields through the Zeeman effect using some strong near-infrared coronal emission lines. However, their integration time was very long ($\sim$70 minutes). Thus, their measurements could not reveal the temporal evolution of the magnetic fields in the highly dynamic corona. Also their observed regions are strong-field active regions. It is very difficult to measure the coronal magnetic fields in the weak-field quiet-Sun regions, e.g., the cavities. 

Fortunately, linear polarization measurements of some coronal forbidden lines can reveal the direction of coronal magnetic fields in the plane of sky, although linear polarization is not sensitive to the magnetic field strength \citep{Charvin1965}. The Coronal Multichannel Polarimeter (CoMP) observes complete polarization states and Doppler velocities of the forbidden lines Fe XIII 10747 {\AA} \& 10798 {\AA} \citep{Tomczyk2008}, which has revealed valuable information on the coronal magnetic fields. For instance, different magnetic field structures often produce different polarization signatures. Through the polarization measurement of CoMP, one can identify possible magnetic field morphology in different regions of the corona \citep[e.g.,][]{Rachmeler2013}. The Doppler shift measurement has also provided significant insight into the magnetic structures of coronal cavities. For instance, \citet{Bak2016} found that a large fraction of cavities possess nested rings of line-of-sight velocity, strongly indicating a line-of-sight oriented and axial magnetic field within the cavities. In addition, the Doppler shift measurements have also revealed significant insight into the propagation of coronal mass ejections \citep{Tian2013} and coronal Alfv\'enic waves \citep{Tomczyk2007,Tomczyk2009,McIntosh2012,Threlfall2013,Liu2014,Morton2015}. Observations of the waves could also provide information on the coronal magnetic field through coronal seismology.  

Magnetic flux ropes are often involved in magnetic field models of cavities \citep[e.g.,][]{priest1989, low1995, gibson1998, van2004, fan2010}. In recent years forward calculations using the toolset of FORWARD \citep{gibson2016} have been often used to help distinguish between different magnetic field models by comparing the calculated and observed linear polarization. For instance, \citet{dove2011} performed FORWARD calculation for a spheromak-type magnetic flux rope in the model of \citet{gibson1998}, and found a ring-shaped structure in the linear polarization that is consistent with CoMP observation of a cavity. With CoMP observations, \citet{Bak2013} found that the ``lagomorphic" signature in linear polarization is common in quiescent prominence cavities. Then they performed FORWARD calculation for an arched cylindrical flux rope, which was taken from the three-dimensional magnetohydrodynamic (MHD) simulation of \citet{fan2010}, and found that the calculated linear polarization also reveals such a lagomorphic structure. In most of these previous investigations, the magnetic field models are given analytically or taken from MHD simulations. The first attempt of applying the FORWARD calculation to magnetic flux-rope models constructed using observed photospheric magnetograms was made by \citet{jibben2016}. These authors also found a lagomorphic structure in the calculated linear polarization, which is consistent with the linear polarization of the associated cavity observed by CoMP. However, the Doppler shift measurement of CoMP does not reveal any nested rings of line-of-sight velocity in this cavity. Obviously, FORWARD calculations for magnetic field structures constructed using the observed magnetograms are still highly desired to improve our understanding of the magnetic environment of the prominence-cavity system.

In this paper, we present an investigation of a cavity observed above the western limb of the Sun during the total solar eclipse on 2017 August 21. We construct a series of magnetic field models based on the magnetograms taken one week before the totality with the Helioseismic and Magnetic Imager \citep[HMI,][]{Schou2012} onboard the Solar Dynamics Observatory (SDO). We then perform FORWARD calculations for these models, and the calculated linear polarization signals are compared with CoMP observations. The model field lines are also compared with the coronal structures seen in images taken by the Atmospheric Imaging Assembly (AIA) \citep{Lemen2012} on board SDO and the coronal red line images taken during our eclipse expedition. These comparisons suggest a highly twisted flux rope within the cavity. The Doppler shift measurement of CoMP also supports this scenario. We also examine the thermal properties of the cavity using the ratio of Fe XIII 10747 {\AA} and Fe X 6374 {\AA}.

\section{Observations and data reduction} \label{sec:obser}
A total solar eclipse occurred across the continental United States on 2017 August 21. Among several professional eclipse observing teams, there is one consisting of solar physicists from Yunnan Observatories (YNO) of Chinese Academy of Sciences, Peking University (PKU), and Sichuan University of Science and Engineering. This eclipse expedition team have performed successful scientific observations of the corona at Dallas, Oregon, located at N$44^{\circ}$55' and W$123^{\circ}$18'. The totality started at 17:16:57 UT and lasted for approximately 1 minute and 57 seconds at this site. A brief description of the telescopes and scientific objectives of this eclipse expedition team can be found in \cite{Tian2017}.

\begin{figure*} 
\centering {\includegraphics[width=\textwidth]{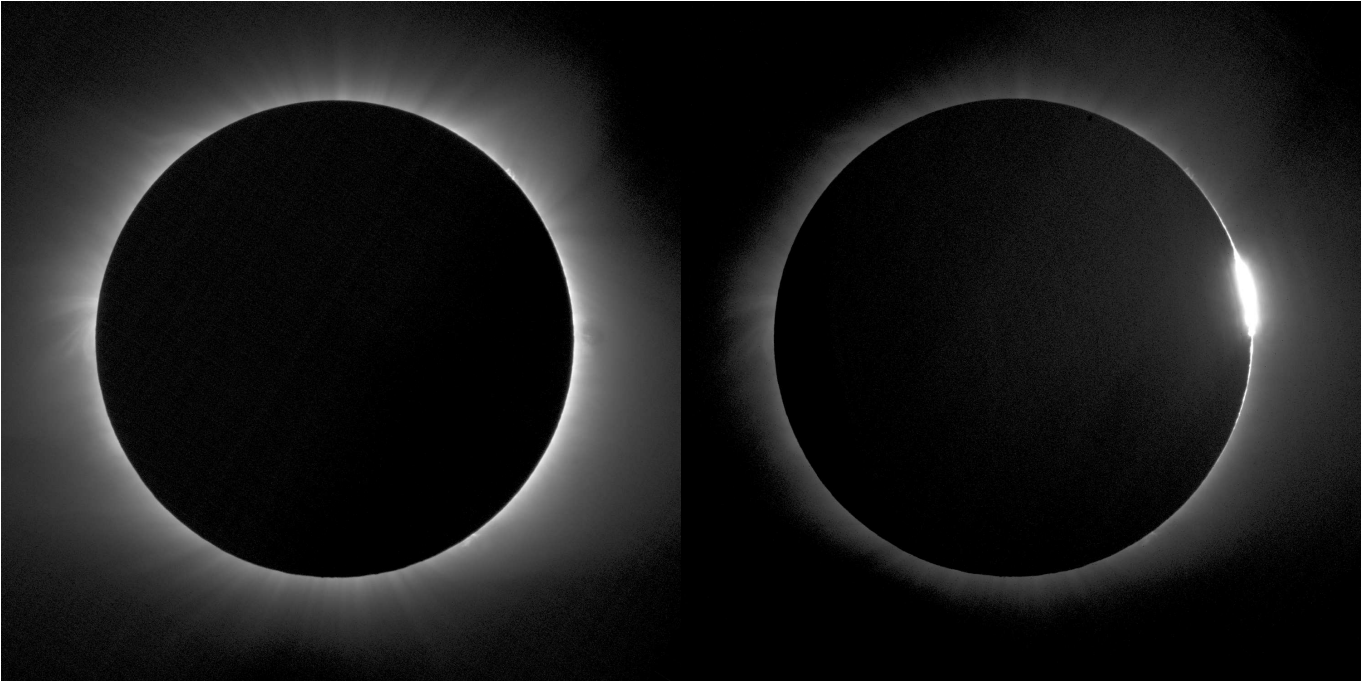}} 
\caption{Images of the coronal red line (Fe X 6374 {\AA}) taken by CRIT at 17:18:41 UT (left panel) and 17:18:56 UT (right panel). The left image shows a cavity above the west limb and the right one shows the Baily's beads.} \label{cavity}
\end{figure*}

Here we use images taken by two telescopes of this eclipse expedition team: PKU$^{\prime}$s Coronal Red line Imaging Telescope (CRIT) and YNO$^{\prime}$s Green Line Imaging Polarimetric Telescope (GLIPT). The images of CRIT were taken through a 5 {\AA} FWHM bandpass filter, centered at the coronal red line Fe X 6374 {\AA}. A LXC-250M camera manufactured by Baumer Group was used for the imaging experiment, which provided data in 10 byte format. The camera has a 5120${\times}$5120 CMOS array with a pixel size of 4.5 ${\mu}$m, coupled with a 15.2 cm aperture refracting telescope with a focal ratio of f/10. The spatial pixel size is 0.76 arcsec. 
The flat field and dark current were assessed after the totality. All the red line images are reduced by subtracting the corresponding dark current and dividing by the normalized flat field. Then all the images are rotated to make the vertical axis aligned in the solar north-south direction. Figure ~\ref{cavity} shows two reduced coronal red line images taken by CRIT during the totality at 17:18:41 UT and 17:18:56 UT, respectively. Since the emission of Fe X 6374 {\AA} is very weak far from the disk center, only part of the full field of view (FOV) is shown. The left image shows a clear cavity structure above the west limb, around the equator. The right image reveals the Baily's beads observed when the totality was just over.

The GLIPT took images of the polarization states of the coronal green line Fe XIV 5303 {\AA} in a wide bandpass (100  {\AA}). This telescope is a refractive telescope with an 8 cm aperture.
A series of flat field and dark current images for each polarization states were taken after the totality. Once the dark currents are subtracted and the flat fields are corrected, the Stokes I images of the coronal green line are obtained. These images are then rotated to make the vertical axis aligned in the north-south direction.

We use the line of sight magnetograms taken by the SDO/HMI instrument to construct magnetic field models. These magnetograms were taken with a cadence of 45 seconds and a pixel size of $\sim$0.5 arcsec. The full-disk SDO/AIA images taken in the passbands of 304 {\AA}, 171 {\AA} and 211 {\AA} are also used in our study. These images were taken with a pixel size of $\sim$0.6 arcsec and a cadence of 12 seconds.

\begin{figure*}
\centering {\includegraphics[width=\textwidth]{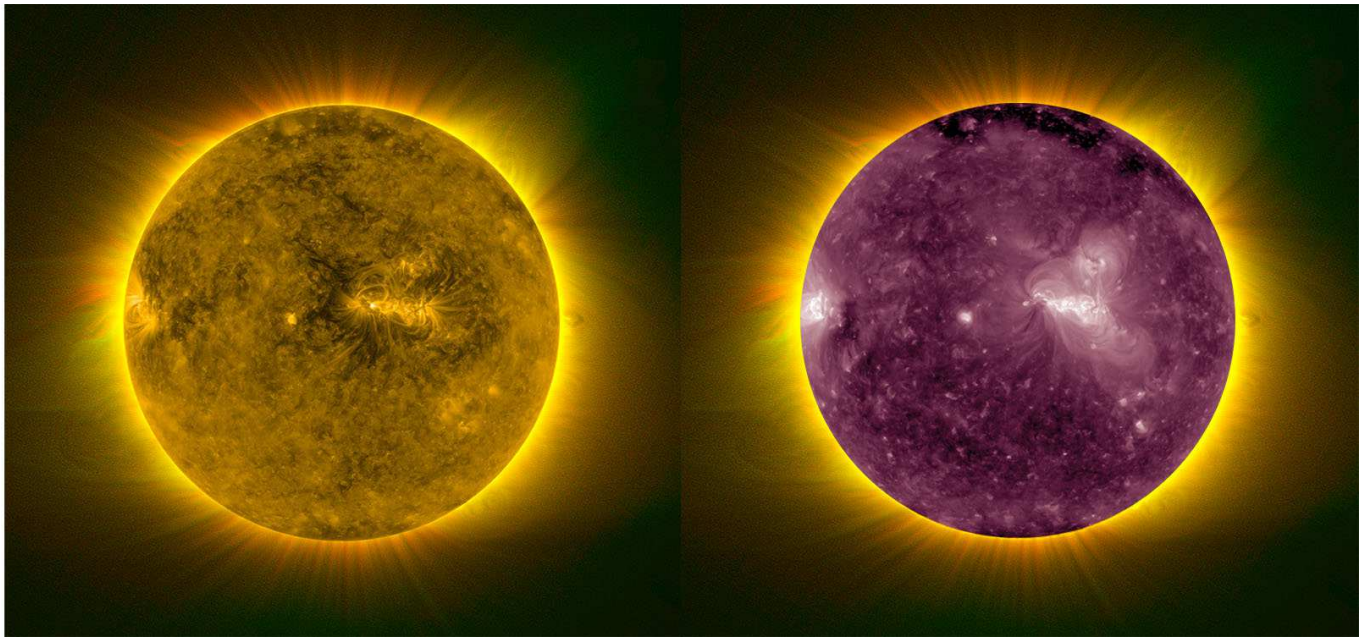}} 
\caption{Composite images of the solar corona. In each panel, the solar disk part is an image taken by SDO/AIA (left: 171 {\AA} passband; right: 211 {\AA} passband). The off-limb part in both panels are the same, showing a composite and enhanced image of the coronal red line and green line. These images were all taken at 17:18 UT. } \label{enhan-img}
\end{figure*}

We have processed both the images of coronal red line and green line using the Noise Adaptive Fuzzy Equalization (NAFE) method developed by Druckm$\ddot{u}$ller \citep{miloslav2013} to enhance fine structures in the corona. Figure ~\ref{enhan-img} presents two composite coronal images. The off-limb parts of both images are the blends of enhanced red line and green line images. The red line images taken at 17:18:41 UT are shown in red, and the green line images are shown in green. The disk parts are images taken by the AIA 171 {\AA} (left panel) and 211 {\AA} (right panel) at 17:18 UT, respectively. It is known that a combination of red line and green line images allows a temperature diagnostics, as under ionization equilibrium the red and green lines are emitted from the plasmas with a temperature of about one million and two millions Kelvin, respectively \citep[e.g.,][]{Habbal2010,Habbal2011}. Figure ~\ref{enhan-img} clearly reveals a lower temperature in polar coronal holes and a higher temperature in active regions. We also find that the cavity above the west limb is much more prominent in the processed/enhanced images.

The CoMP instrument also performed coronal observations during the totality. The CoMP images have a pixel size of $\sim$4.5 arcsec. The FOV of CoMP is 1.04 $-$ 1.4 R$_{\odot}$. We use the Fe XIII 10747 {\AA} data sampled at three spectral locations of 10745.0 {\AA}, 10746.2 {\AA} and 10747.4 {\AA} during the totality. Sequential images of different polarization states (Stokes I, Q, and U) were obtained at each of these three spectral locations with a cadence of 30 seconds. Using the analytical Gaussian fit introduced by \citet{Tian2013}, we can deduce the intensity and Doppler velocity of the Fe XIII line. By treating the line-averaged Stokes Q and U as the Stokes Q and U of the emission line, the total linear polarization L can be calculated as $\sqrt{Q^2+U^2}$.

\section{Magnetic field modeling} \label{sec:model}
To understand the magnetic structure of the cavity, we construct magnetic field models using the flux rope insertion method developed by van Ballegooijen \citep{van2004}. We briefly introduce the method below, for detailed descriptions please refer to \citet{bobra2008} and \citet{su2009, su2011}.

\begin{figure}[ht!]
\plotone{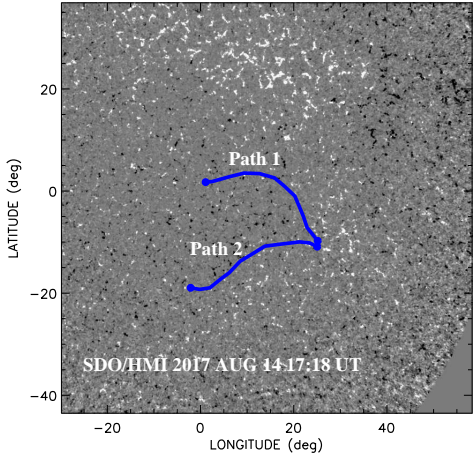}
\caption{The longitude--latitude images of the radial component of the photospheric magnetic field by SDO/HMI in the HIRES region at 17:18 UT on 2017 August 14. The blue curves with circles at the two ends refer to the paths along which we insert flux ropes. 
\label{model-path}}
\end{figure}

At first, a potential field model (Model 2) is computed from the high-resolution (HIRES) and global magnetograms observed by SDO/HMI. Since the prominence is observed near the west limb, we have to use a magnetogram taken several days before the prominence/cavity observed on 2017 August 21. Similar methodology has also been used in our previous studies \citep{su2012, su2015}. The lower boundary condition for the HIRES region is derived from the photospheric line-of-sight magnetograms obtained at 17:18 UT on August 14. The longitude-latitude map of the radial component of the magnetic field in the HIRES region is presented in Figure ~\ref{model-path}. The HIRES computational domain extends about 88$^{\circ}$ in longitude, 80.3$^{\circ}$ in latitude, and up to 2 $\rm R_\odot$ from the Sun center. The models use variable grid spacing to achieve high spatial resolution in the lower corona (i.e., $0.002 R_{\sun}$) while covering a large coronal volume in and around the target region. Two paths marked with blue curves are selected according to the locations of the observed filaments. Path 1 represents the filament corresponding to our target cavity observed on the limb, and another quiescent filament is represented by path 2. Next we modify the potential field to create cavities in the region above the selected paths, then insert two thin flux bundles (representing the axial flux $\Phi_{\rm axi}$ of the flux rope) into the cavities. Circular loops are added around the flux bundle to represent the poloidal flux $F_{\rm pol}$ of the flux rope. The resulted magnetic fields are not in force-free equilibrium. We then use the magneto-frictional relaxation to drive the field towards a force-free state \citep{van2000, yang1986}. 

\begin{figure*}
\centering {\includegraphics[width=\textwidth]{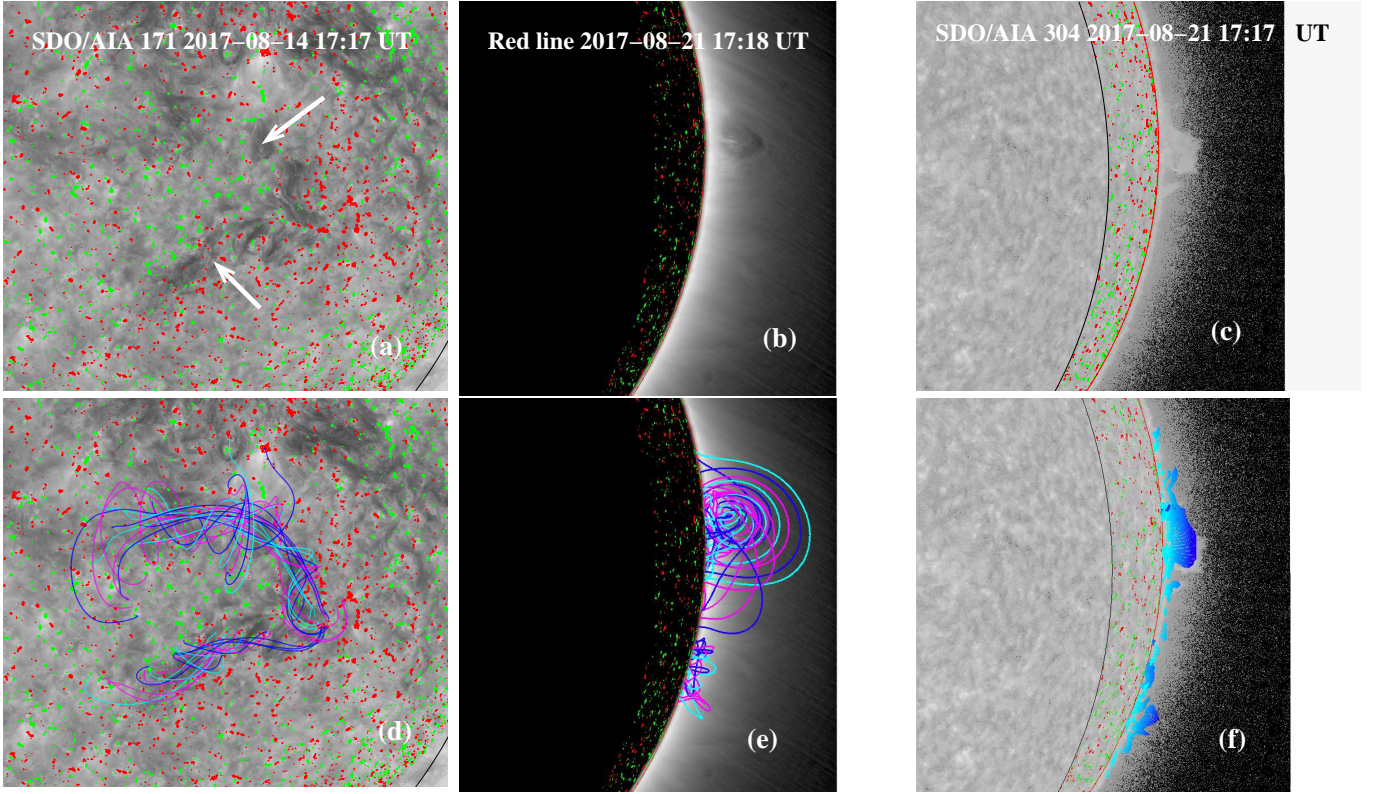}}
\caption{Comparison between observations and the best-fit magnetic field model (Model 1) after 30000-iteration relaxations. The top panels from left to right, present images in 171 {\AA} by SDO/AIA, red line by CRIT, and 304 {\AA} by SDO/AIA. Corresponding images overlaid with selected field lines (color lines) or dips (blue features in panel f)  from the best-fit model are presented in the bottom panels. The locations of dips in the field lines refer to the sites where the field lines are locally horizontal and curved upward (light blue for low-lying dips, darker blue at larger heights). The red and green contours in panels (a) and (d) represent positive and negative magnetic polarities taken by SDO/HMI. }
\label{model-obs}
\end{figure*}

We construct a series of magnetic field models by varying the axial and poloidal fluxes of the inserted flux ropes. One of the best-fit models (Model 1) after 30000-iteration relaxations is presented in Figure ~\ref{model-obs}. The initial inserted poloidal fluxes are $-10^{10}$ and 0 Mx cm$^{-1}$,  and the axial fluxes are 2$\times10^{20}$ and $10^{20}$ Mx for paths 1 and 2, respectively. SDO/AIA 171 {\AA} image is presented on the top left panel, which shows that this region has two quiescent filament channels marked by white arrows. The cavity and prominence structure is shown on the top middle and right panels. The bottom row shows the same images overlaid with selected field lines from the best-fit model. The shape, location and size of the model field lines match the observed filament channel and cavity well. The dips of the model field lines also closely match the observed limb prominence. The northern filament channel and cavity correspond to a twisted flux rope, while the southern quiescent filament channel is consistent with sheared-arcade structure.

The twist angle of the flux rope  can be estimated as $\Phi = 2 \pi F_{pol} L / \Phi_{axi} $ \citep{2009ApJ...703.1766S}. ``L" refers to the length of the flux rope, which is estimated by the length of a randomly selected field line in the center of the flux rope \citep{su2011}. This length is about $4.5\times10^{10}$ cm for the northern flux rope. The poloidal flux of the northern flux rope decreases from $-10^{10}$ Mx cm$^{-1}$ at the beginning to $-7\times10^{9}$ Mx cm$^{-1}$ after 30000-iteration relaxations. Accordingly, the twist angle of the northern flux rope is 3.1$\pi$ after 30000-iteration relaxations. The northern flux rope is kink stable,  since the twist angle (3.1$\pi$) is below the critical twist (3.5$\pi$) for kink instability in the numerical modeling in more idealized configurations \citep[e.g., ][]{2003ApJ...589L.105F, 2004ApJ...609.1123F, 2004ApJ...617..600G, 2004A&A...413L..23K, 2004A&A...413L..27T}, although it exceeds the critical twist (2.5$\pi$) derived by \citet{hood1981} for a line-tying force-free magnetic flux rope. The twist angle (3.1$\pi$) of the flux rope is consistent with the most probable value (2$\pi$ $-$ 4$\pi$) of the total twist angle of interplanetary flux ropes observed at 1 AU according to recent statistical study by \citet{wang2016}.  

\section{FORWARD calculation} \label{sec:forward}

Linear polarization of the Fe XIII 10747 {\AA} line is sensitive to the direction of magnetic field in the plane-of-sky (POS). When the magnetic field is solely in the POS, or perpendicular to the line-of-sight (LOS), the magnitude of linear polarization reaches the maximum. The signal of linear polarization vanishes when the magnetic field is aligned with the LOS or the inclination angle of the magnetic field from local vertical equals $54.7^{\circ}$. The later is known as the Van Vleck effect \citep{van1925}. Thus the depletion of the magnitude of linear polarization is very useful for examining the magnetic field direction. 

\begin{figure*}
\centering {\includegraphics[width=\textwidth]{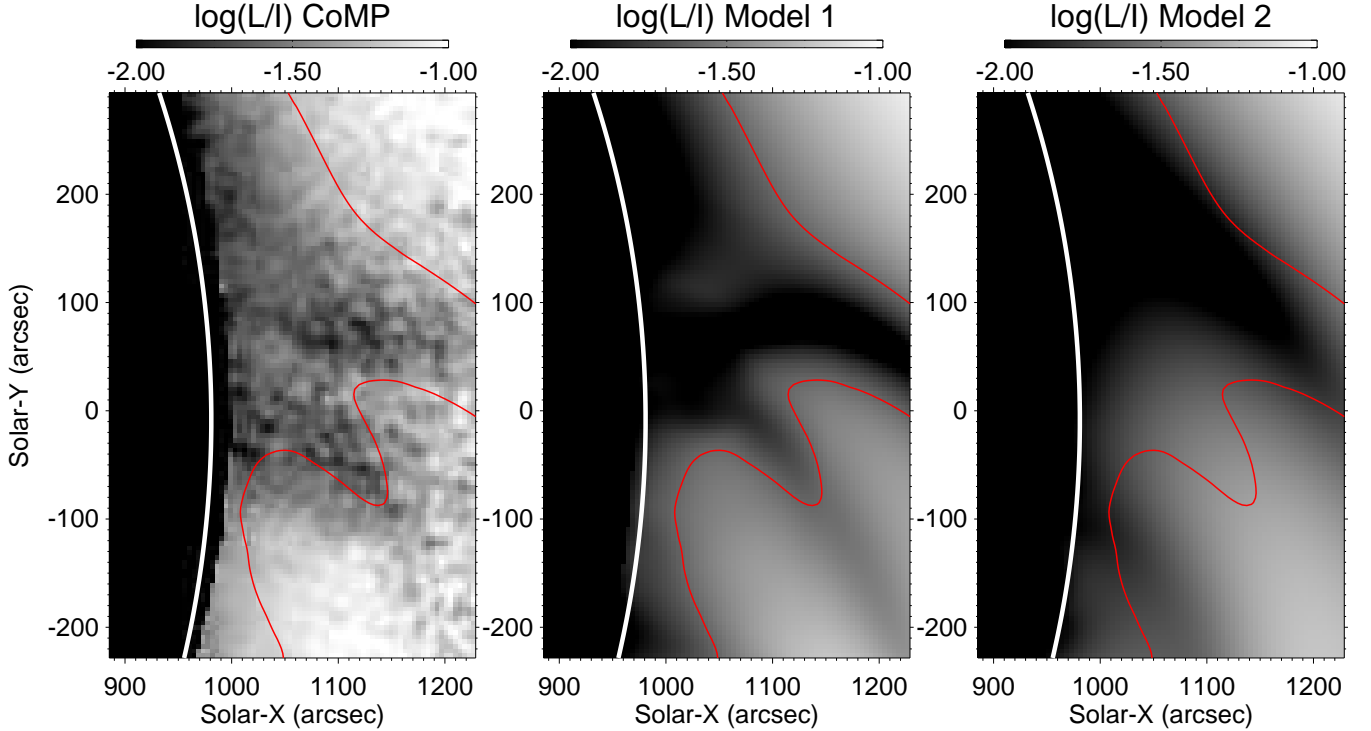}} 
\caption{Left: Degree of linear polarization (L/I, log scale) from CoMP observations of the Fe XIII 10747 {\AA} line at 17:21 UT on 2017 August 21. Middle: L/I calculated for Model 1 (best-fit model). Right: L/I calculated for Model 2 (potential field model). The contours represent locations where the log$_{10}$(L/I) equals $-$1.6 in Model 1. Model 1 nicely reproduces the lagomorphic structure in linear polarization observed by CoMP. The white line in each panel marks the edge of the moon at 17:18 UT.} 
\label{model-result}
\end{figure*}

To compare the constructed magnetic field models with CoMP observations, we need to calculate the linear polarization from our models. The FORWARD IDL package \citep{gibson2016} available in SolarSoft can be used to calculate the Stokes vector produced along a given LOS for magnetic field models at the Fe XIII 10747 {\AA} line. The FORWARD calculations are based on the location and local plasma parameters including the magnetic field, temperature, density, and height above the solar surface. In our models, we assume an isothermal corona with a temperature of 1.5 MK. The density decreases with height exponentially to balance the gravity. Under these assumptions and using the constructed magnetic field models, we have calculated Stokes I, Q, U, V using FORWARD. We then obtain the linear polarization degree L/I, where $L=\sqrt{Q^2+U^2}$. 

We perform FORWARD calculations for different magnetic field models we constructed, and compare the modeling results with COMP observations. As mentioned above, the magnetic field lines of Model 1 match the structures in AIA images and coronal red line images best. We find that this model also matches the CoMP observations of polarization best. Figure ~\ref{model-result} compares L/I (logarithmic scale) measured by CoMP with the theoretical L/I calculated from Model 1 and Model 2 (potential field model). It is obvious that Model 1 reproduces the lagomorphic structure in linear polarization observed by CoMP, while the potential field model reproduces only one ear. This result implies that the magnetic structure of the cavity is a twisted flux rope, which produces the lagomorphic structure in linear polarization. A similar result was previously obtained by \citet{Bak2013}. However, the flux rope used in their calculation is not constructed from observations. The mismatch between the potential field model and CoMP observation indicates that this cavity-prominence system is highly non-potential. A lot of free energy should be stored in such a system, which could provide the energy for possible eruptions. This cavity appears to be different from the cavity with a weakly twisted flux rope in \citet{jibben2016}. Their calculations show that the potential field model and the best-fit model have very similar signatures in linear polarization. 

\section{Line-of-sight velocity observed by CoMP}\label{sec:dop}

\begin{figure*}
\centering {\includegraphics[width=0.8\textwidth]{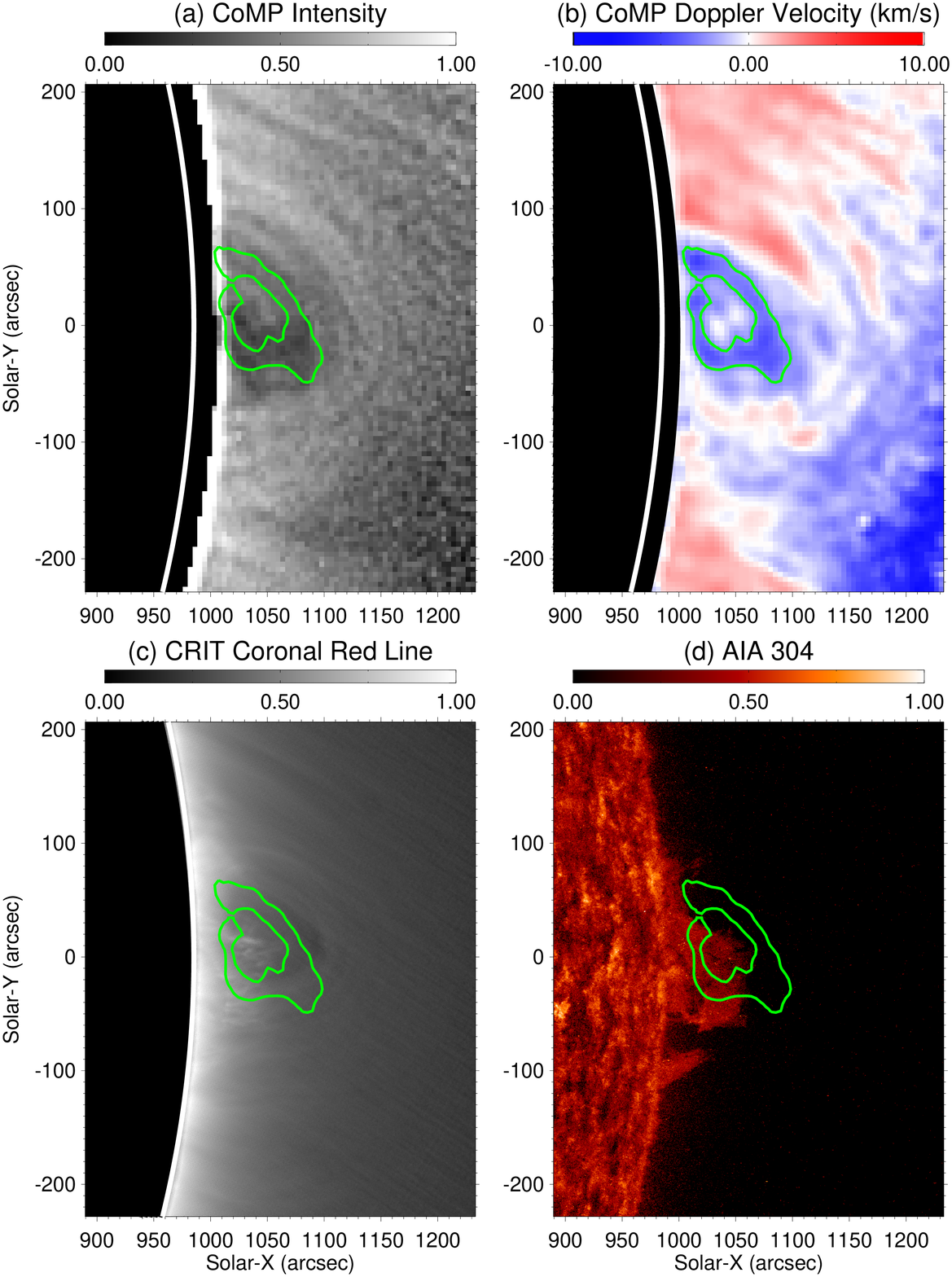}} 
\caption{(a) The cavity seen in the enhanced line center intensity of Fe XIII 10747 {\AA} observed by CoMP. (b) Doppler velocity of Fe XIII 10747 {\AA} observed by CoMP. (c) The cavity seen in the NAFE-processed coronal red line image observed by CRIT. (d) The cavity-associated prominence observed in the AIA 304 {\AA} passband. The FOV is the same as that of Figure ~\ref{model-result}. Images shown in the panel (a) and (b) were taken at 17:14 UT, and  (c) and (d) at 17:18 UT on 2017 August 21. The ring-shaped structure in the Doppler velocity image is outlined by green contours with a level of $-$2.6 km/s. The white line in panels (a)--(c) marks the edge of the moon at 17:18 UT.}
\label{doppler}
\end{figure*}

Besides linear polarization, CoMP can also measure Doppler shift of the forbidden line Fe XIII 10747 {\AA}, which reveals the LOS velocity component of plasma flows. Figure ~\ref{doppler} presents the line center intensity and Doppler velocity obtained by using the analytical Gaussian fit \citep{Tian2013}. The intensity image has been enhanced using the radial filtering method developed by \citet{forland2013}. A coronal red line image taken by CRIT and the simultaneously taken AIA 304 {\AA} image are also present in this figure. The AIA 211 {\AA} images taken during the totality are used to co-align the AIA and CoMP observations. The red line image has been processed by using the NAFE method.

Arch-shaped structures have been previously reported from measurements of LOS velocities within cavities \citep{Schmit2009}. In our CoMP observation we see an obvious ring-shaped structure in the Doppler velocity image. The ring is located exactly in the cavity that is visible in the Fe XIII 10747 {\AA} and red line images. This ring-shaped Doppler pattern appears to be similar to the ring-shaped structures reported by \citet{Bak2016}, indicating hot plasma flows along a helical magnetic structure. Such a scenario is consistent with the magnetic field configuration of a flux rope with its axis aligned in the LOS. So the Doppler shift measurement also supports the interpretation of the cavity as a flux rope. It is worth mentioning that such a Doppler pattern was not found in the cavity studied by \citet{jibben2016}, which may be related to the fact that the magnetic fields within their cavity are only weakly twisted. Future statistical investigations may need to be performed to examine whether such ring-shaped Doppler shift structures are present only in highly twisted flux ropes.

The cavity-related prominence observed in the AIA 304 {\AA} passband appears to be spatially offset from the center of the ring. Also the top of the prominence is not located below the largest Doppler velocity, which is different from the results in previous statistical study of \citet{Bak2016}. However, the prominence still lies at the bottom of the cavity and at the dips of the magnetic field lines in Model 1. 

\section{\textbf{Temperature structure of the cavity}}\label{sec:temp}

The intensity ratio (R$_I$) of Fe XIII 10747 {\AA} and Fe X 6374 {\AA} is sensitive to the electron temperature. Thus, information about the temperature of the cavity can be inferred from this ratio, if we assume an isothermal plasma along the line of sight. It is worth mentioning that in the case of ionization equilibrium this method is valid only in the temperature range of $\sim$$10^{6}$ K --$10^{6.3}$ K, where both the Fe XIII 10747 {\AA} and Fe X 6374 {\AA} lines are present. The relationship between the calibrated intensities (I$_{XIII}$, I$_{X}$) and the observed counts (I$_{XIII\_obs}$, I$_{X\_obs}$) for both lines can be expressed in the following way:

$$ I_{XIII}  = K_{XIII} \times I_{XIII\_obs} $$
$$ I_{X}  = K_{X} \times I_{X\_obs} $$
$$ R_I = \frac{I_{XIII}}{I_{X}} = \frac{K_{XIII}}{K_{X}} \times \frac{I_{XIII\_obs}}{I_{X\_obs}} = \frac{K_{XIII}}{K_{X}} \times R$$
where $R=\frac{I_{XIII\_obs}}{I_{X\_obs}}$ is the observed intensity ratio of Fe XIII 10747 {\AA} and Fe X 6374 {\AA}, and K$_{XIII}$ and K$_{X}$ are the calibration factors for Fe XIII 10747 {\AA} and Fe X 6374 {\AA}, respectively. Unfortunately K$_{X}$ could not be precisely determined since the exposure time was not accurately recorded during the totality. As a result, we could not determine the absolute values of the temperature at different locations. However, since this intensity ratio monotonically increases with the temperature, we can still examine where the temperature is higher or lower by comparing the values of \textit{R} at different locations.

\begin{figure*}
\centering {\includegraphics[width=\textwidth]{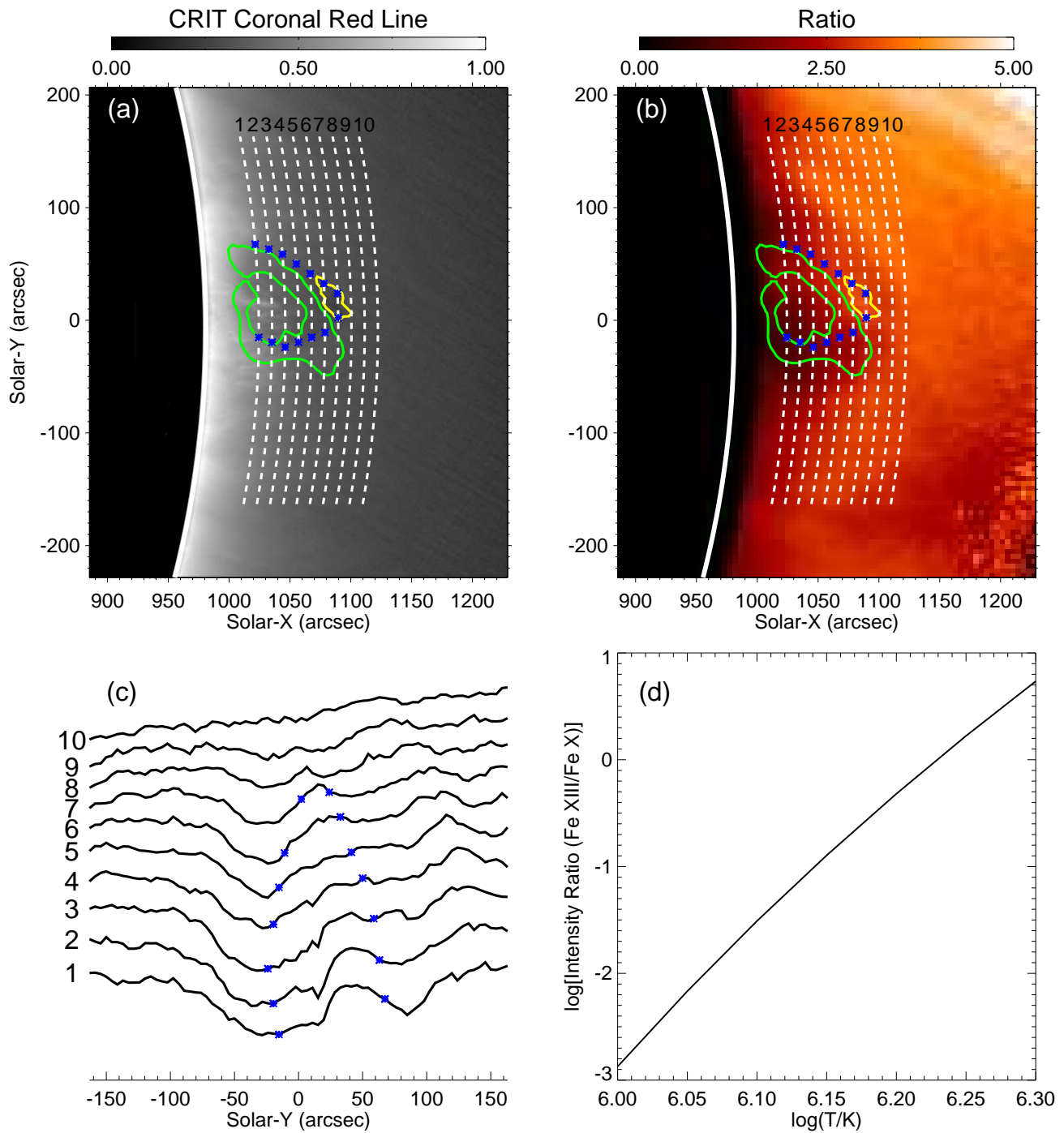}}
\caption{(a): The cavity seen in the NAFE-processed coronal red line image. (b): The ratio (\textit{R}) of the observed intensities of Fe XIII 10747 {\AA} and Fe X 6374 {\AA}, calculated from the unprocessed images. The green contours are the same as in Figure ~\ref{doppler}. The yellow contours represent locations where \textit{R} equals 3 within the cavity. The white solid lines mark the edge of the moon at 17:18 UT. (c): The values of \textit{R} along the ten white arcs shown in the upper panels. The boundary of the cavity is marked by the blue asterisks in panels (a) \& (b), and indicated by two asterisks on the curves of arcs 1--7 in panel (c).  (d): The dependence of the intensity ratio ($R_I$) of Fe XIII 10747 {\AA} and Fe X 6374 {\AA} on the electron temperature, as calculated from CHIANTI v7.1. }
\label{temp_map}
\end{figure*}

Using the unprocessed images of Fe XIII 10747 {\AA} and Fe X 6374 {\AA} observed by CoMP and CRIT, respectively, we have obtained the spatial distribution of \textit{R} in the corona. Figure ~\ref{temp_map}(a) \& (b) present the NAFE-processed red line image and the image of \textit{R}, respectively. Figure ~\ref{temp_map}(c) shows the variation of \textit{R} along ten arcs that are placed at different heights above the limb. These arcs are parallel to the limb. The theoretical relationship between the Fe XIII 10747 {\AA}/Fe X 6374 {\AA} intensity ratio ($R_I$) and electron temperature, as calculated from CHIANTI v7.1 \citep{Landi2013} under the assumption of ionization equilibrium, is presented in Figure ~\ref{temp_map}(d). Here we can see that a larger $R_I$ indicates a higher temperature. Since $R_I$ is proportional to \textit{R}, we can conclude that the temperature is higher when \textit{R} is larger. 

We find that \textit{R} is nonuniform within the cavity, suggesting different temperatures at different parts of the cavity. The southern part of the cavity appears to reveal smaller \textit{R} values, indicating lower temperatures there. The \textit{R} values, and thus the temperatures, are higher in the northern part of the cavity. We also find a high-\textit{R} region at the top of the cavity (the yellow contours in Figure ~\ref{temp_map}(a) \& (b)), which just sits above the flux rope marked by the Doppler contours. It is possible that most cold plasmas are constrained at lower heights and there are mainly hot materials above the flux rope. Another possible explanation is that the plasmas there are heated through magnetic reconnection at the separator between the magnetic flux rope within the cavity and the ambient magnetic fields, or dynamical friction$/$hyperdiffusion along the magnetic flux surface \citep{van2008}. The temperature structure of this cavity appears to be more complex than that of \citet{Reeves2012}, where an obvious temperature increase was found in the core of the cavity. It is also possible that at some pixels there are multiple emitting sources with different temperatures along the line of sight, which can affect our temperature diagnostics.

\section{Summary}\label{sec:sum}
We present results from an investigation of a coronal cavity observed above the western limb in the coronal red line and green line by two telescopes operated respectively by Peking University and Yunnan Observatories during the total solar eclipse on 2017 August 21. To understand the magnetic field structure of the cavity, a series of magnetic field models are constructed based on the magnetograms taken by SDO/HMI. We obtain the linear polarization for different models through FORWARD calculations, and compare the model outputs with CoMP observations of the linear polarization.

The magnetic field lines in the best-fit model match the coronal structures in the red line images and AIA 171 {\AA} images. The dips of the model field lines match the limb prominence observed in the AIA 304 {\AA} passband. From this model we have also reproduced the ``lagomorphic" signature in the linear polarization of the Fe XIII 10747 {\AA} line that is observed by CoMP. In addition, we find a ring-shaped structure in the line-of-sight velocity in the cavity, which implies hot plasma flows along a helical magnetic structure in the cavity. These results indicate that the magnetic structure of the cavity is a highly twisted flux rope. We have also investigated the thermal properties of the cavity using the intensity ratio of Fe XIII 10747 {\AA} and Fe X 6374 {\AA}, and found a nonuniform distribution of temperature within the cavity.

The twist angle of the flux rope in the best-fit model is 3.1$\pi$. This value exceeds the critical twist (2.5$\pi$) derived by \citet{hood1981} for a line-tying force-free magnetic flux rope, but is below the critical twist (3.5$\pi$) for kink instability given by numerical modeling. On the other hand, this twist angle is consistent with the most probable value (2$\pi$ $-$ 4$\pi$) of the total twist angle of interplanetary flux ropes observed at 1 AU. It is unclear whether the observed flux rope in the cavity will eventually erupt or not.

\begin{acknowledgements}
The eclipse observation and analysis of the data are supported by NSFC grants 11790304 (11790300), 11373065, 11527804, 11473071, 41574166 and 41231069, the Recruitment Program of Global Experts of China, the One Hundred Talent Program of CAS, and the Max Planck Partner Group program. We thank Sarah Gibson for assisting us to set up FORWARD, and Steven Tomczyk and Joan Burkepile for providing the CoMP data. SDO is a mission of NASA's Living With a Star (LWS) Program. CoMP is an instrument funded by NCAR/HAO and NCAR is sponsored by the National Science Foundation. CHIANTI is a collaborative project involving George Mason University, the University of Michigan and the University of Cambridge.
\end{acknowledgements}


\begin{thebibliography}{}

\bibitem[B\c ak-St\c e\' slicka et al.(2013)]{Bak2013} B\c ak-St\c e\' slicka, U., Gibson, S. E., Fan, Y., et al. \ 2013, \apjl, 770, L28

\bibitem[B\c ak-St\c e\' slicka et al.(2016)]{Bak2016} B\c ak-St\c e\' slicka, U., Gibson, S. E., and Chmielewska, E.\ 2016, Front. Astron. Space Sci. 3:7

\bibitem[Bobra et al.(2008)]{bobra2008} Bobra, M.~G., van Ballegooijen, A.~A., \& DeLuca, E.~E.\ 2008, \apj, 672, 1209-1220

\bibitem[Charvin(1965)]{Charvin1965} Charvin, P. \ 1965, AnAp, 28, 877

\bibitem[Dove et al.(2011)]{dove2011} Dove J. B., Gibson S. E., Rachmeler L. A., Tomczyk S. \& Judge P.\ 2011, \apjl, 731:L1

\bibitem[Druckm$\ddot{u}$ller M.(2013)]{miloslav2013} Druckm$\ddot{u}$ller, M.\ 2013, \apjs, 207, 25

\bibitem[Fan(2010)]{fan2010} Fan, Y. \ 2010, \apj, 719, 728

\bibitem[Fan \& Gibson(2003)]{2003ApJ...589L.105F} Fan, Y., \& Gibson, S.~E.\ 2003, \apjl, 589, L105

\bibitem[Fan \& Gibson(2004)]{2004ApJ...609.1123F} Fan, Y., \& Gibson, S.~E.\ 2004, \apj, 609, 1123 

\bibitem[Forland et al.(2013)]{forland2013} Forland, B. C., Gibson, S. E., Dove, J. B., Rachmeler, L. A, and Fan, Y. \ 2013, Sol. Phys. 288, 603

\bibitem[Fuller \& Gibson(2009)]{fuller2009} Fuller, J., and Gibson, S. E. \ 2009, \apj, 700, 1205

\bibitem[Gibson \& Low(1998)]{gibson1998} Gibson, S. E. \& Low, B. C. \ 1998, \apj, 493, 460

\bibitem[Gibson et al.(2004)]{2004ApJ...617..600G} Gibson, S.~E., Fan, Y., Mandrini, C., Fisher, G., \& Demoulin, P.\ 2004, \apj, 617, 600 

\bibitem[Gibson et al.(2006)]{gibson2006} Gibson, S. E., Foster, D., Burkepile, J., de Toma, G., \& , A. S. \ 2006, \apj, 641, 590

\bibitem[Gibson et al.(2010)]{gibson2010} Gibson, S. E., Kucera, T. A., Rastawicki, D., et al. \ 2010, \apj, 724, 1133

\bibitem[Gibson et al.(2016)]{gibson2016} Gibson, S. E., Kucera, T. A., White, S. M., et al. \ 2016, Front. Astron. Space Sci. 3, 8

\bibitem[Habbal et al.(2010)]{Habbal2010} Habbal, S. R., Druckm$\ddot{u}$ller, M., Morgan, H., et al. 2010, ApJ, 719, 1362

\bibitem[Habbal et al.(2011)]{Habbal2011} Habbal, S. R., Druckm$\ddot{u}$ller, M., Morgan, H., et al. 2011, ApJ, 734, 120

\bibitem[Harvey(1969)]{Harvey1969} Harvey, J. W. 1969, Ph.D. thesis, Univ. Colorado

\bibitem[Heinzel et al.(2008)]{Heinzel2008} Heinzel P., Schmieder B., F\' arn\' ik, F., \ 2008, \apj, 686, 1383

\bibitem[Hood \& Priest(1981)]{hood1981} Hood, A. W., \& Priest, E. R.\ 1981, Geophys. Astrophys. Fluid Dyn., 17, 297

\bibitem[Hudson et al.(1999)]{hudson1999} Hudson, H. S., Acton, L. W., Harvey, K. A., \& McKenzie, D. M. \ 1999, \apj, 513, 83

\bibitem[Hudson \& Schwenn et al.(2000)]{hudson2000} Hudson, H. S. \& Schwenn, R. \ 2000, Adv. Space Res., 25, 1859

\bibitem[Illing \& Hundhausen et al.(1986)]{illing1986} Illing, R. M., \& Hundhausen, J. R. \ 1986, \jgr,  91, 10951

\bibitem[Jibben et al.(2016)]{jibben2016} Jibben, P. R., Reeves, K. K., \& Su, Y. \ 2016, Front. Astron. Space Sci. 3:10

\bibitem[Kliem et al.(2004)]{2004A&A...413L..23K} Kliem, B., Titov, V.~S., T{\"o}r{\"o}k, T.\ 2004, \aap, 413, L27

\bibitem[Kuhn(1995)]{Kuhn1995} Kuhn, J. R. 1995, in National Solar Observatory/Sacramento Peak Summer Workshop, IR Tools for Solar Astrophysics: What's Next?, ed. J. R. Kuhn \& M. J. Penn (Singapore: World Scientific), 89

\bibitem[Landi et al.(2013)]{Landi2013} Landi, E., Young, P. R., Dere, K. P., Del Zanna, G., \& Mason, H. E. 2013, ApJ, 763, 86

\bibitem[Lemen et al.(2012)]{Lemen2012} Lemen, J. R., Title, A. M., Akin, D. J., et al.\ 2012, Sol. Phys. 275, 17

\bibitem[Lin et al.(2000)]{lin2000} Lin, H., Penn, M. J., \& Tomczyk, S. \ 2000, \apjl, 541, L83

\bibitem[Lin et al.(2004)]{lin2004} Lin, H., Kuhn, J. R., and Coulter, R. \ 2004, \apjl, 613, L177

\bibitem[Liu et al.(2014)]{Liu2014} Liu, J., McIntosh, S. W., De Moortel, I., Threlfall, J., Bethge, C. \ 2014, \apj, 797, 7

\bibitem[Low \& Hundhausen(1995)]{low1995} Low, B. C., \& Hundhausen, J. R. \ 1995, \apj, 443, 818

\bibitem[Marqu\'e(2004)]{Marque2004} Marqu\'e, C. 2004, \apj, 602, 1037

\bibitem[Marqu\'e et al.(2002)]{Marque2002} Marqu\'e, C., Lantos, P., \& Delaboudinere, J.-P. 2002, \aap, 387, 317

\bibitem[McIntosh \& De Pontieu(2012)]{McIntosh2012} McIntosh, S. W.,  De Pontieu, B. \ 2012,  \apj, 761, 138

\bibitem[Morton et al.(2015)]{Morton2015} Morton, R. J., Tomczyk, S., Pinto, R. 2015, Nature Communications, 6, 7813

\bibitem[Priest et al.(1989)]{priest1989} Priest, E. R., Hood, A. W., \& Anzer, U.\ 1989, \apjl, 344, 1010:1025

\bibitem[R\'egnier et al.(2011)]{Regnier2011} R\'egnier, S., Walsh, R.W., \& Alexander, C.E. 2011, \aap, 533, L1

\bibitem[Rachmeler et al.(2013)]{Rachmeler2013} Rachmeler, L. A., Gibson, S. E., Dove, J. B., DeVore, C. R., Fan, Y. 2013, Solar Physics, 288, 617

\bibitem[Reeves et al.(2012)]{Reeves2012} Reeves K. K., Gibson S. E., Kucera T. A., Hudson H. S. \& Kano R. \ 2012, \apj, 746, 146

\bibitem[Savcheva \& van Ballegooijen(2009)]{2009ApJ...703.1766S} Savcheva, A., \& van Ballegooijen, A.\ 2009, \apj, 703, 1766 

\bibitem[Schou et al.(2012)]{Schou2012} Schou, J., Scherrer, P. H., Bush, R. I., et al.\ 2012, Sol. Phys. 275, 229

\bibitem[Schmit et al.(2009)]{Schmit2009} Schmit, D. J.,Gibson, S. E., Tomczyk, S., et al. \& Tripathi, D. 2009, ApJ, 700, L96

\bibitem[Su et al.(2009)]{su2009} Su, Y., van Ballegooijen, A., Schmieder, B., et al.\ 2009, \apj, 704, 341 

\bibitem[Su et al.(2011)]{su2011} Su, Y., Surges, V., van Ballegooijen, A., DeLuca, E., \& Golub, L.\ 2011, \apj, 734, 53 

\bibitem[Su \& van Ballegooijen(2012)]{su2012} Su, Y., \& van Ballegooijen, A.\ 2012, \apj, 757, 168 

\bibitem[Su et al.(2015)]{su2015} Su, Y., van Ballegooijen, A., McCauley, P., et al.\ 2015, \apj, 807, 144 

\bibitem[Tandberg-Hanssen(1995)]{Tandberghanssen1995} Tandberg-Hanssen, E. \ 1995, The Nature of Solar Prominences, 2nd Edn. Dordrecht: Kluwer.

\bibitem[Threlfall et al.(2013)]{Threlfall2013} Threlfall, J., De Moortel, I., McIntosh, S. W., \& Bethge, C. 2013, A\&A, 556, A124

\bibitem[Tian et al.(2013)]{Tian2013}	Tian, H., Tomczyk, S., McIntosh, S. W., et al.\ 2013, Sol. Phys, 288, 637

\bibitem[Tian et al.(2017)]{Tian2017}	Tian, H., Qu, Z. Q., Chen, Y. J., et al. \ 2017, Earth and Planetary Physics, 1, 68

\bibitem[Tomczyk et al.(2007)]{Tomczyk2007} Tomczyk, S., McIntosh S. W., Keil S. L., et al. 2007, Science, 317, 1192

\bibitem[Tomczyk et al.(2008)]{Tomczyk2008} Tomczyk, S., Card, G. L., Darnell, T., et al. \ 2008, Sol. Phys, 247(2), 411

\bibitem[Tomczyk \& McIntosh(2009)]{Tomczyk2009} Tomczyk, S., McIntosh, S. W.\ 2009,  \apj, 697, 1384 

\bibitem[T{\"o}r{\"o}k et al.(2004)]{2004A&A...413L..27T} T{\"o}r{\"o}k, T., Kliem, B., \& Titov, V.~S.\ 2004, \aap, 413, L27 

\bibitem[van Ballegooijen(2004)]{van2004} van Ballegooijen, A.~A.\ 2004, \apj, 612, 519 

\bibitem[van Ballegooijen et al.(2000)]{van2000} van Ballegooijen, A.~A., Priest, E.~R., \& Mackay, D.~H.\ 2000, \apj, 539, 983 

\bibitem[van Ballegooijen \& Cranmer (2008)]{van2008} van Ballegooijen, A. A., \& Cranmer, S. R. \ 2008, \apj, 682, 644

\bibitem[van Vleck(1925)]{van1925} van Vleck, J.H.\ 1925, Proc. Natl. Acad. Sci. USA 11, 612, 618.

\bibitem[Wang et al.(2016)]{wang2016} Wang, Y., Zhuang, B., Hu, Q., et al. \ 2016, \jgr, 121, 9316

\bibitem[Wesley, W. H.(1927)]{Wesley1927} Wesley, W. H. \ 1927, Memoirs of the British Astronomical Association, Vol. 64, Appendix

\bibitem[Yang et al.(1986)]{yang1986} Yang, W.~H., Sturrock, P.~A., \& Antiochos, S.~K.\ 1986, \apj, 309, 383

\end{thebibliography}
\end{document}